# Effect of tube radius on the adsorption of chlorothalonil on single-walled carbon and boron nitride nanotubes surfaces: A theoretical study for environmental remediation

**Francisco Gleidson de S. Ferreira[1], Caio V. C. Ribeiro da Silva[2], Silvete Guerini[3], Kleuton A. L. Lima[4–5], Douglas Soares Galvão[5], José Milton Elias de Matos[1], Alexandre Araujo de Souza[1],***

*Corresponding author.
E-mail address: alesouza@ufpi.edu.br (A.A. Souza)

[1] Graduate Program in Chemistry, Federal University of Piauí, 64049-550, Teresina, PI, Brazil
[2] Department of Chemistry, Federal University of São Carlos, 13565-905, São Carlos, SP, Brazil
[3] Department of Physics, Federal University of Maranhão, 65080-805, São Luís, MA, Brazil
[4] Institute of Physics Gleb Wataghin, State University of Campinas, 13083-859, Campinas, SP, Brazil
[5] Institute of Physics Gleb Wataghin, State University of Campinas, 13083-859, Campinas, SP, Brazil

**ABSTRACT**

The interaction of the pollutant chlorothalonil (CLT) with single-walled carbon nanotubes (CNTs) and boron nitride nanotubes (BNNTs) was investigated using density functional theory (DFT) calculations with the SIESTA software. Structural, energetic, and electronic properties were analyzed to characterize the effects of the interaction on the nanotube surfaces. The adsorption energy increases with increasing nanotube radius, which is attributed to enhanced π–π stacking interactions. The results indicate that both CNTs and BNNTs can physically adsorb CLT, but BNNTs emerge as more promising candidates for CLT detection and removal in contaminated environments. The nanotube radius strongly affects all electronic and energetic properties. Molecular dynamics simulations performed in GROMACS confirmed the stability of CLT adsorbed on CNT and BNNT surfaces in aqueous solution.

**1. Introduction**

Chlorothalonil (CLT) (2,4,5,6-tetrachlorobenzene-1,3-dicarbonitrile) is a fungicide commonly used on a variety of fruits and vegetables to control the growth of ascomycetes and fungi imperfecti [1],[2]. However, when used in excess, it can pose serious risks to human and animal health as it is highly toxic to mammals and aquatic animals [3]. In addition, CLT is also used as an additive biocide activator in some types of antifouling paints and as a wood and paint preservative [4]. It has a half-life of 60 days [5] and is classified as a human carcinogen by the United States Environmental Protection Agency (USEPA) [6].

The removal of CLT is important because it is relatively stable and adheres well to plants, resulting in higher levels of residues in the environment [7]. Repeated spraying of CLT leads to its release into the environment, and its presence in surface and groundwater is of concern [8].

Wang *et al*. found that the rate of CLT degradation decreases with increasing concentration as this compound is physically adsorbed to the soil [9]. Lopes et al. showed that CLT impairs the physiognomy of red blood cells and fish fertility [10]. Among the various techniques for removing pollutants from the environment, adsorption is characterized by its efficiency with high removal rates even at very low concentrations [11–13], and nanotubes are promising materials for this process due to their large surface area [14].

Since their discovery, carbon nanotubes (CNTs) have been intensively researched due to their great potential for the detection and removal of toxic molecules in aqueous or gaseous media as well as for other technological applications such as the development of supercapacitors [15], hydrogen storage, the fabrication of sieves, membranes, separation filters, purification systems [16,17], X-ray sources, optoelectronic devices [18], drug delivery, antibiotic application, tissue generation [19] and chemotherapy-resistant sensors [20].

On the other hand, boron nitride nanotubes (BNNTs) exhibit interesting physical and chemical properties, similar to carbon nanotubes. Currently, BN materials are used in insulating materials, optical diodes operating in the ultraviolet range, UV detectors, cosmetics, sensors, cutting tools, substrates for electronic devices, and lubricants [21].

Unlike CNTs, their electronic properties are independent of chirality [22], i.e. all BNNTs are semiconducting materials, while CNTs can behave as conductors or semiconductors depending on chirality [23]. Studies have shown that BNNTs have high oxidation resistance, hardness, high mechanical strength, high thermal stability, and conductivity with heat resistance [24] compared to CNTs, making them one of the most favorable materials for applications such as molecular sensors in the inorganic and organic fields.

Theoretical studies show that nanotubes are good candidates for the removal of pollutants. The strength of adsorption generally depends on the orientation of the molecule with respect to the nanotube surface as well as the nature of both the nanotube and the molecule. Serejo et al. have shown in a theoretical study that FeCl$_3$ and CrO$_3$ molecules are chemically adsorbed on the outer surface of BNNT [25]. Bouhara and Hammoutène have investigated the interaction of glyphosate on the CNT and BNNT surfaces and have shown that a physical adsorption process takes place [26]. Another theoretical study shows the use of CNTs as adsorbents for ethinylestradiol and estrone molecules, and in both cases this occurs by physical adsorption [27] Faramarzi et al [28] investigated the interaction of the anticancer drug fluorouracil with CNTs and BNNTs and found that the interaction of fluorouracil with CNTs occurs by physical adsorption and with BNNTs by chemical adsorption. CNTs and BNNTs were studied in detail, and Zhao et al [29] showed through theoretical calculations that both (10,0) and (17,0) CNTs have no difference in adsorption energy with $H_2O$ and $NO_2$ molecules. Duverger et al [30] studied on the interaction of BNNTs with anticancer molecules.

DFT calculations show that CNTs and BNNTs are good candidates for putrescine and cadaverine sensors [31]. Recently, single-walled CNTs were reported to be a good adsorbent in conjunction with gas chromatography for the determination and removal of CLT. In another study, CNTs were used to determine organochlorine pesticides (chlorothalonil and pentachloronitrobenzene) in water using dispersive solid-phase extraction (DSPE), followed by gas chromatography (GC). The optimal adsorption conditions were determined by analyzing the effect of adsorbent dosage, adsorption time, eluent type and volume, and elution time. Under the optimal conditions, good linearity was obtained at concentrations from 10 to 400 μg $L^{-1}$ with correlation coefficients ranging from 0.9991 to 0.9986. The limits of detection (LOD) for the two organochlorine pesticides were 0.025 and 0.049 μg $L^{-1}$, and the limits of quantification (LOQ) were 0.080 and 0.156 μg $L^{-1}$, respectively. The accuracy of the proposed method was evaluated by measuring the recovery of the spiked samples, which ranged from 82.5% to 110.5% at spiking levels of 0.5–10 μg $L^{-1}$ with relative standard deviations lower than 5.6% (n = 6). This method was successfully applied to determine the target analytes in canal water, drinking water, and water taken from the inlets and outlets of a wastewater treatment plant. The results demonstrate that the developed method has great potential for determining the two organochlorine pesticides in water samples. The results were satisfactory in terms of sensitivity and recovery [32]. However, the role of the nanotube radius in the interaction of this molecule with the CNT has not yet been sufficiently clarified. With this in mind, this article presents a theoretical study on the effect of tube radius on the interaction of chlorothalonil (CLT) with single-walled carbon nanotubes (CNTs) with the following chiralities: (8,0), (10,0) and (12,0). We also evaluate the potential of single-walled boron nitride nanotubes (BNNTs) with the same chiralities detecting and removing this pollutant compared to carbon nanotubes, using density functional theory (DFT).

## 2. Computational methodology

A theoretical-computational study was performed to determine how the nanotube radius affects the structural, energetic and electronic properties of CNTs and BNNTs interacting with chlorothalonil in different configurations and orientations. For this purpose, we combined periodic and non-periodic models with classical, semi-empirical and *ab initio* (DFT) methods.

### 2.1 Periodic models

The periodic models were built with 128, 160 and 192 atoms in the supercell (1:1 ratio of B:N in the case of BNNT) for the (8,0), (10,0) and (12,0) nanotubes, respectively. With this number of atoms, both carbon nanotubes and boron nitride nanotubes have an approximate length of 17 Å, which is larger than twice the average CLT length, ensuring a considerable separation between the periodic images. And, to prevent the nanotube and its periodic images from interacting, a lateral spacing of more than 40 Å was used.

### 2.2 Non-periodic models

The non-periodic models were created by replicating each periodic model twice along the nanotube axis. The atoms were then cut to ensure both edges were equal and terminated in a complete hexagonal ring. Finally, hydrogen atoms were added to complete the valence shell of the edge atoms, resulting in nanotubes with 280 atoms and a length of approximately 27 Å. A simulation box with 50 Å edges was constructed around each nanotube and then solvated with water using a built-in GROMACS tool [33,34]. Subsequently, employing another GROMACS built-in tool, 5 CLT molecules were randomly introduced by replacing some water molecules, resulting in 3994 water molecules in each simulation box.

### 2.3 Electronic structure calculations

The non-periodic systems were also used to perform the initial screening of the preferential position of the CLT molecule on the nanotube, using the software xTB 6.3.3 with the Hamiltonian GFN1-xTB[35]. We decided to first use the semi-empirical tight-binding DFT (xTB) method to reduce computational cost. The initial positions were: (I) Molecule's plane parallel to the tube surface, (II) Molecule's plane perpendicular to the tube surface, Molecules inside the tubes (III) perpendicular and (IV) parallel to the tube radius. The most stable configuration for each nanotube was then optimized using the SIESTA [36] code, in a periodic supercell as described above.

These geometry optimizations were carried out through first-principles calculations based on density functional theory (DFT), using a double-zeta plus polarization (DZP) basis set to represent the valence of the wave function, the generalized gradient approximation (GGA)[37] parameterized by Lee-Murray-Kong-Lundqvist-Langreth (vdW-DF2) [38] to describe the exchange and correlation term, and the Troullier-Martins pseudopotential to describe the interaction between the core and valence electrons [39]. The supercell method was used for the periodic boundary condition and a cutoff radius of 300 Ry to represent the charge density. The Brillouin zone was represented by 18 k-points along the Γ-X direction as proposed by Monkhorst-Pack[40]. The atoms were fully relaxed until the force on each atomic coordinate was less than 0.05 eVÅ$^{-1}$.

The energetic properties were calculated from the most stable configurations, and the adsorption energy ($E_{ads}$) with basis set superposition error (BSSE) correction was calculated using equation (1).

$$E_{ads} = E_{NT+M} - E_{NT} - E_M + E_{BSSE} \qquad (1)$$

where $E_{NT+M}$ is the energy of the interacting system (nanotube + CLT), $E_{NT}$ is the energy of the pristine nanotube, $E_M$ is the energy of the isolated CLT molecule, and $E_{BSSE}$ is the basis set superposition error correction. The electronic properties were analyzed using the projected density of states (PDOS). The charge transfer ($CT$) was calculated for each system using the Bader charge. The band gap ($E_g$) was measured using the band structure as the difference between the top of the valence band and the bottom of the conduction band. The electronic sensitivity ($E_{NT+M}$) and the work function sensitivity ($\Delta\emptyset$) for the interaction between the CLT and the nanotube was calculated using equation (2),

$$\Delta X\% = \left(\frac{X_{(NT+M)} - X_{(NT)}}{X_{(NT)}}\right) \times 100\% \qquad (2)$$

where $X_{(NT)}$ and $X_{(NT+M)}$ represent a given electronic property of the nanotubes before and after the interaction, respectively. This property can be either the band gap (Eg) or work function (ø).

Transition state theory was used to determine the recovery time ($\tau$) as a function of adsorption energy ($E_{ads}$) through equation (3),

$$\tau = v_0^{-1} e^{-\left(\frac{E_{ads}}{kT}\right)} \qquad (3)$$

where $v_0$ is the attempt frequency ($10^{12}$ s$^{-1}$), $k$ is the Boltzmann constant and $T$ is the thermodynamic temperature [41]. The electrical conductivity ratio (σ/σ₀) was estimated using the following expression (equation 4):

$$\sigma/\sigma_0 = e^{(-\Delta E_g/2k)} \qquad (4)$$

where $E_g$ is the band gap of the system, $k$ is the Boltzmann constant and $T$ is the thermodynamic temperature [42].

The chemical reactivity descriptors (chemical potential, hardness and electrophilicity index) were determined using the vertical ionization potential (IP) and electron affinity (EA). These were obtained from separate single-point calculations on the cationic (+1) and anionic (-1) states (the ΔSCF method), respectively, thus avoiding the Koopmans' theorem approximation based on frontier molecular orbital energies.

Charge density difference (Δρ) plots were obtained by subtracting the charge density of the interacting system from the sum of the charge densities of its isolated fragments (nanotube and molecule). Ghost atoms were included in the isolated fragments to ensure an accurate representation of the electron density using the same basis set size.

**2.4 Molecular dynamics simulations**

All molecular dynamics (MD) simulations were performed in the GROMACS suite (version 2020.7) [33,34], using the OPLS-AA [43] force field, and the TIP4P water model. The bonded parameters for the BNNT nanotubes were taken from the work of Rajan and co-workers [44], whereas non-bonded parameters were taken from Hilder et al., who optimized them for BNNT–water interactions. [45]. The bonding parameters used for the B-N-H and N-B-H angles correspond to the B-N-B and N-B-N angles taken from Rajan and coworkers, respectively. The partial charges for these terminal atoms were taken from the CM5 charges [46] for a calculation of the borazine molecule performed with xtb 6.3.3 software using the GFN1-xTB Hamiltonian [35]. The N-H (B-H) charges were -0.63786 and 0.40178 (0.23004 and 0.00604) for the N (B) and H atoms, respectively. Although the parameters for these terminal bonds were not well parameterized, our MD simulations were performed to investigate the interaction of the two nanotubes with CLT molecules in water solution. Since both nanotubes are very long, the parameters for these terminal bonds will not significantly affect our results. Also, all bonds involving hydrogen atoms were constrained using the 4th order LINCS algorithm[47].

The MD protocol was performed in the following steps: (1) energy minimization using the steepest descent algorithm; (2) 1 ns of a NVT molecular dynamics simulation using the v-rescale algorithm (T = 298.15 K, $\tau T$ = 0.1 ps), followed by a NPT MD simulation with the same simulation time and temperature but using the Berendsen barostat [48] for pressure equalization (P = 1bar, $\tau P$ = 1 ps), and finally (3) 100 ns of a NPT production step with the same parameters (T = 298.15 K, $\tau T$ = 0.1 ps) using the Parrinello-Rahman barostat [49] (P = 1bar, $\tau P$ = 5 ps). All MD were performed with a timestep of 2 fs.

## 3. Results and discussions
### 3.1 Structural and adsorption energetic properties

After many calculations with different orientations of the molecule on the nanotube, it was found that the most stable configuration was the one in which the ring of the molecule was parallel to the surface of the nanotube (Fig. 1). In this configuration, the molecule was located at an initial distance of 2.00 Å between the center of the molecular ring and the surface of the nanotube. After geometry optimization, this distance was increased to values between 3.38 and 3.45 Å, depending on the system (Table 1). It is important to note that lower distances were obtained on the larger nanotubes, which can indicate an increase in the interaction between the CLT and both nanotubes, as the radius increases.

**Fig. 1** Optimized geometries of the (a) (8,0) CNT, (b) (10,0) CNT, (c) (12,0) CNT, (8,0) BNNT, (10,0) BNNT and (12,0) BNNT interacting with the chlorothalonil molecule, showed in side (left) and frontal (right) view. The yellow, green, light blue and gray spheres represent the C, Cl, N and B atoms, respectively.

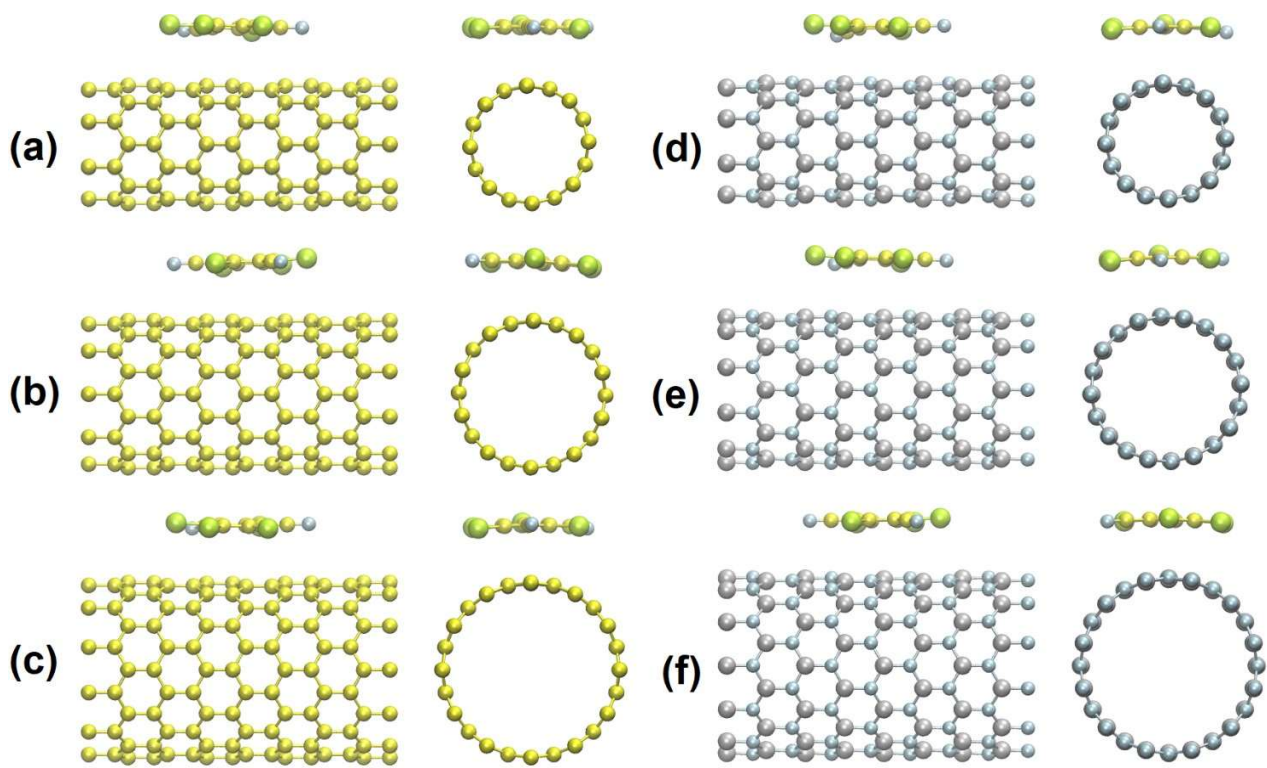

The distances between the CLT and the tube surfaces are very far from a conventional chemical bond length between these elements, indicating that it is not a chemical but a moderate physical adsorption. This can be confirmed by the adsorption energy ($E_{ads}$) values (Table 1) calculated using equation 1, whose values tend to decrease with the radius of the tubes [50,51]. This correlation also confirms the idea of better interaction between these nanotubes and CLT with increasing radius of the nanotubes. We will investigate this effect in more detail when we analyze the energy states of these systems.

*Table 1:* Adsorption energy ($E_{ads}$), recovery times (τ) and average distance (R) between the CLT aromatic ring and each nanotube surface for each interacting system. The negative sign in charge transfer means that the CLT receives charge from the nanotube atoms.

| Systems | $E_{ads}$ (eV) | τ (s) | R (Å) |
|---|---|---|---|
| (8,0) CNT + CLT | -0.70 | 0.68 | 3.43 |
| (10,0) CNT + CLT | -0.76 | 7.00 | 3.40 |
| (12,0) CNT + CLT | -1.13 | 1.27x10$^7$ | 3.38 |
| (8,0) BNNT + CLT | -0.55 | 0.001 | 3.40 |
| (10,0) BNNT + CLT | -0.66 | 0.14 | 3.41 |
| (12,0) BNNT + CLT | -0.79 | 22.72 | 3.45 |

Both the (8,0), (10,0), and (12,0) BNNTs demonstrated lower adsorption energies and short recovery times (Table 1), which are desirable characteristics for reversible sensing [42]. The short recovery time values indicate that CLT is spontaneously adsorbed and desorbed from the nanotube surface frequently, making the BNNT reusable. However, the same cannot be inferred for the CNTs, because as their radius increases, both the adsorption energy and the recovery time significantly increase. Notably, the (12,0) CNT, the largest carbon nanotube analyzed in this study, exhibits a very high recovery time. Since for adsorption energies between -0.34 eV and -0.79 eV the recovery time ranges from 0.5 μs to 16 s and for an adsorption energy of -1.00 eV the recovery time corresponds to 12 h, [41], the (12,0) CNT is not suitable for this application. It is therefore expected that CNTs with radii larger than (12,0) would exhibit even longer recovery times. In this sense,

BNNTs are a more suitable alternative than CNTs for detecting and removing CLT from contaminated environments.

### 3.2 Electronic and work function sensing properties

As a result of a physical adsorption process, we do not expect significant changes in the electronic properties of the nanotube, and this indeed remains true for the CNTs, where there were very small changes of 0.01~0.08 eV in the band gap (Table 2), resulting in a negligible electronic sensitivity ($\Delta Eg$). However, for the BNNTs, significant changes in the band gap were observed, yielding high electronic sensitivities. Due to the significantly band gap reduction, the electrical conductivity (Table 2) also changes, which can be used as a parameter for detecting the CLT in the environment. The charge transfer analysis (Table 2) using the Bader charge analysis (CT) is fully consistent with the expected physics, confirming the electron donor tendency of the nanotubes in all cases, and that the BNNT systems exhibit lower charge transfer values than CNT systems. Furthermore, the CT revealed a clear trend, as the diameter of the CNTs increases, the charge transfer also increases. While in BNNTs, there is no significant charge variation, which is consistent with their weaker interaction [52].

While the Bader analysis quantifies the charge transfer, the charge density difference ($\Delta\rho$) plots provide a detailed spatial understanding of this process, reconciling the observed trends with the underlying electronic interactions. The $\Delta\rho$ plots (Fig. 2) confirm that the primary charge flow originates from the nanotube region beneath the molecule's aromatic ring (a blue isosurface of charge depletion) towards the electronegative sites on the adsorbate, such as the aromatic ring itself, the chlorine atoms, and the cyano group (red isosurfaces of charge accumulation). In the case of CNTs, this charge donation process is the dominant feature, leading to a straightforward and greater charge transfer.

**Fig. 2** Charge density difference ($\Delta\rho$) plots for the interacting systems with (a) (8,0) CNT, (b) (10,0) CNT, (c) (12,0) CNT, (d) (8,0) BNNT, (e) (10,0) BNNT and (f) (12,0) BNNT. The yellow, green, light blue and gray spheres represent the C, Cl, N and B atoms, respectively. The red and blue isosurfaces correspond to regions of charge accumulation and depletion, respectively. All isosurfaces were plotted to enclose 80% of the total rearranged charge.

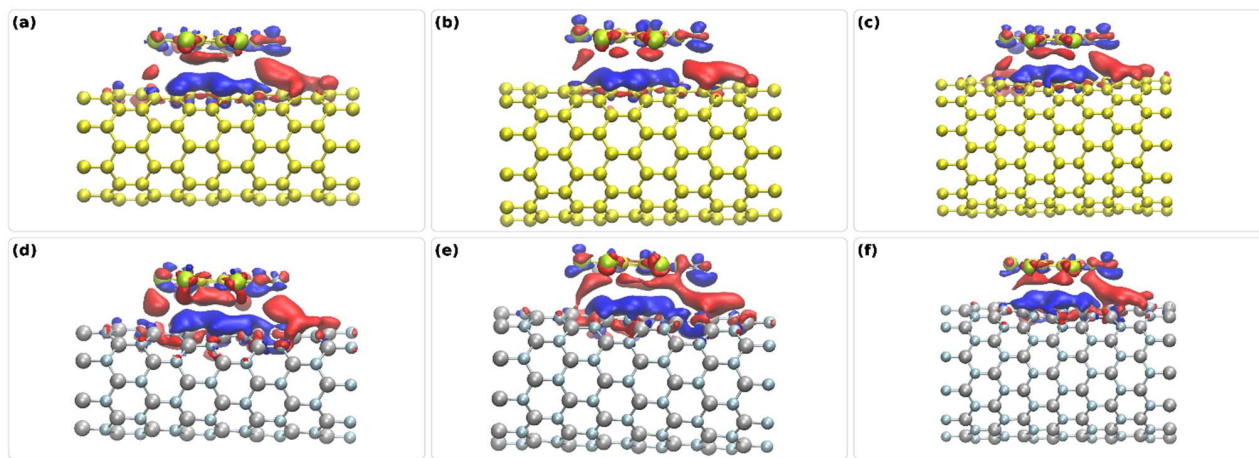

Conversely, the BNNT systems, despite showing a smaller charge transfer in the Bader analysis, exhibit a more complex and polarized electronic rearrangement. The $\Delta\rho$ plots (see Fig. 2 (d)-(f)) reveal that while BNNTs also display a significant, often more voluminous, region of charge depletion under the aromatic ring, this effect is strongly counteracted by a pronounced charge accumulation on the nanotube surface, particularly beneath the molecule's cyano group. This indicates a more significant bidirectional charge displacement and strong local polarization, likely driven by the intrinsic polarity of the B-N bonds interacting with the molecule's dipole. The presence of these large, competing regions of both charge donation and accumulation on the BNNT surface results in a partial cancellation of the overall charge flow, leading to the smaller charge transfer values quantitatively captured by the Bader method.

A particularly notable feature in all systems is the strong polarization of the cyano group, visualized as a 'sandwich' of charge density. This observation is fully consistent with the electronic nature of CLT, an electronically active molecule known to exhibit significant intramolecular charge transfer. The cyano group, in particular, acts as a strong electron-accepting group and is in conjugation with the aromatic ring [53]. Therefore, the polarization observed in our system after adsorption is consistent with the electronic nature of this functional group, which is clearly exacerbated by the physisorption interaction with the surface.

*Table 2:* Band gap ($E_g$), electronic sensitivity ($\Delta E_g$), work function ($\varnothing$) and work function variation ($\Delta\varnothing$), electrical conductivity ratio ($\sigma/\sigma_0$) and charge transfer calculated through Bader charges (CT) for each interacting system in comparison with the non-interacting systems. All $\sigma$ and $\tau$ values were calculated assuming a temperature of 298.15 K.

| Systems | $E_g$ (eV) | $\Delta E_g$ (%) | $\varnothing$ (eV) | $\Delta\varnothing$ (%) | $\sigma/\sigma_0$ | CT (e) |
|---|---|---|---|---|---|---|
| (8,0) CNT | 0.70 | - | 4.68 | - | - | - |
| (8,0) CNT + CLT | 0.69 | 1.43 | 4.71 | -0.64 | 1.21 | -0.03 |
| (10,0) CNT | 0.60 | - | 4.50 | - | - | - |
| (10,0) CNT + CLT | 0.58 | 3.33 | 4.54 | -0.88 | 1.47 | -0.04 |
| (12,0) CNT | 0.19 | - | 4.47 | - | - | - |
| (12,0) CNT + CLT | 0.18 | 5.26 | 4.49 | -0.45 | 1.21 | -0.05 |
| (8,0) BNNT | 3.64 | - | 4.00 | - | - | - |
| (8,0) BNNT + CLT | 2.19 | 39.84 | 5.34 | -25.09 | $1.8\times10^{12}$ | -0.01 |
| (10,0) BNNT | 4.07 | - | 4.39 | - | - | - |
| (10,0) BNNT + CLT | 2.19 | 46.19 | 5.35 | -17.94 | $7.8\times10^{15}$ | -0.01 |
| (12,0) BNNT | 4.25 | - | 5.25 | - | - | - |
| (12,0) BNNT + CLT | 2.15 | 49.41 | 5.37 | -2.24 | $5.5\times10^{17}$ | -0.01 |

Recently, single-walled CNTs were used experimentally to determine CLT in wastewater samples. CNT were used as a solid phase for gas chromatography and showed good removal rates from the water samples [32]. Although the CNT diameter in their study was larger than the maximum diameter considered in this work, their result agrees with ours, because for CNTs to be used in solid-phase extraction, they must have good interaction with CLT, and as we have previously shown, increasing the nanotube radius also increases the adsorption energy. However, their study only focused on removing CLT in a sample containing CLT. As far as we know, no experimental studies are showing that CNTs can detect CLT, and our results suggest that this is impractical because the change in the electronic properties of CNTs is very small.

Further investigations of CNT and BNNT sensitivity for CLT detection were carried out using the work function. The work function of a semiconductor is the energy required to remove an electron from the Fermi level to a point where it no longer influences the material. Some measuring devices are based on work function changed transistors. They are based on the change in the work function of the material due to adsorption, which changes its field emission properties. Table 2 shows that the work function (calculated as the negative of the Fermi level) remains practically unchanged after the interaction of CLT with the CNT surface, while it is significantly increased after the interaction with the BNNT surface for the (8,0) and (10,0) nanotubes. However, for (12,0) BNNT, the work function is very small, indicating that the increase of the nanotube radius decreases the emitted electron current and the sensitivity to this molecule. From this, we can conclude that the BNNT's radius must be well controlled for this application. In addition, negative values of %$\Delta\varnothing$ indicate a decrease in the work function, i. e., the Fermi level was shifted to the valence band compared to the pristine nanotube, which may be attributed to charge transfer from nanotube to the molecule. This direction of charge transfer is consistent with the Bader charge analysis results (Table 1). Furthermore, this process effectively changes the electronic properties of the nanotube, enhancing its electrical conductivity, which is confirmed by the observed reduction in the system's band gap (Table 2).

In Fig. 3(a)–(f), we have the band structures of the pristine (left panels) and interacting systems (right panels) respectively. For CNTs (see Fig. 3(a)-(c)), we observe that new localized energy levels appear at the lower end of the conduction band due to the molecule interaction. However, these new flat levels do not contribute significantly to the band gap reduction. For the band structure of BNNTs (Fig. 3(d)-(f)), we can observe that the adsorption of the CLT molecule on the nanotube wall gives rise to three energy levels located above the Fermi level, as well as degenerate levels appearing at the top of the valence band, leading to a reduction in the band gap. The energy gap is significantly reduced due to the introduction of defect levels that appear in the band gap region above the Fermi level (Fig. 3(d)-(f)).

**Fig. 3** Representation of the band structures of the non-interacting (left) and interacting systems (right) with (a) (8,0) CNT, (b) (10,0) CNT, (c) (12,0) CNT, (d) (8,0) BNNT, (e) (10,0) BNNT and (f) (12,0) BNNT. The Fermi level was shifted to zero.

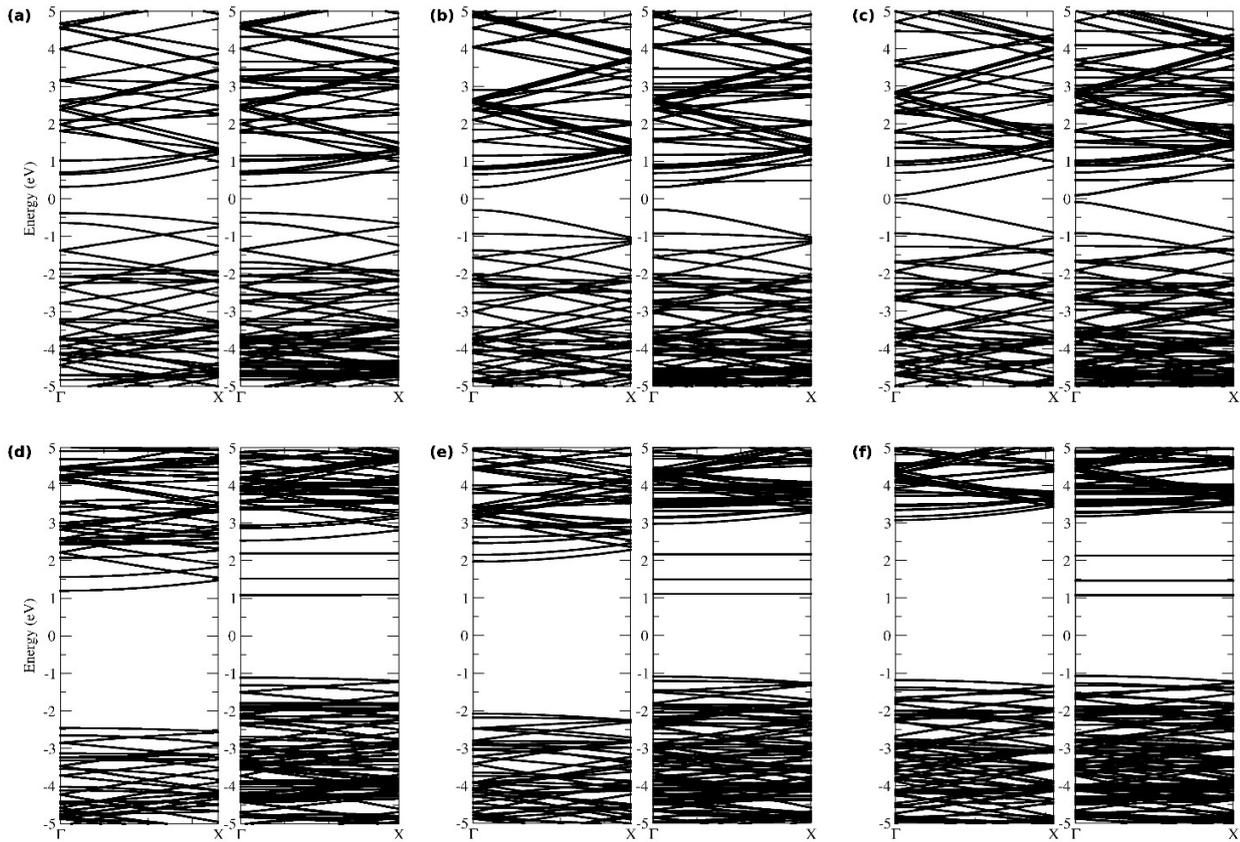

To further investigate the new energy levels that appear in the band structure (see Fig. 3 (a)-(c) right) after the adsorption of the CLT molecule and identify the contribution of each atomic species in the interaction process, we plotted, in Fig. 4, the local density of states (LDOS) for the states at the top of the valence band and bottom of the conduction band, corresponding to 95% probability of finding the electrons. Fig. 4 (a) and (b) show the LDOS for (8,0) CNT plus CLT in the region between -0.39 to -0.27 eV and 0.29 to 0.41 eV, respectively. In Fig. 4 (c), we plotted the LDOS for (10,0) CNT plus CLT in the region between -0.42 and -0.29 eV, while in Fig. 4 (d), we plotted the LDOS for (10,0) CNT plus CLT in the region between 0.35 and 0.50 eV. In Fig. 4 (e) and (f), we plotted the LDOS for (12,0) CNT in the region between -0.28 and -0.20 eV and (12,0) CNT plus CLT in the region between 0.25 and 0.32 eV, respectively. From this analysis, we conclude that the largest contributions to the states at the top of the valence band come from the CNT atoms, with a negligible contribution from the atoms of the CLT molecule (Fig. 4 (a), (c) and (e)). While the contributions to the states at the bottom of the conduction band are from both the CNT atoms and the molecule ((Fig. 4 (b), (d) and (f)). The molecule contribution from these states is less pronounced in (8,0) CNT, and more pronounced in (10,0) and (12,0) CNT.

Fig. 4 (g)-(l) show the LDOS plots for the BNNT systems, corresponding to the band structures in Fig. 3 (d)-(f) on the right. Fig. 4 (g) and Fig. 4 (h) correspond to the LDOS of the (8,0) BNNT plus CLT in the regions between -1.09 to -0.97 eV and 0.99 eV to 1.11 eV, respectively. Fig. 4 (i) to (j) correspond the LDOS for the (10,0) BNNT interacting with the CLT molecule in the regions between -1.09 to -0.97 eV and 0.99 to 1.12 eV. And lastly, Fig. 4 (k) to (l) illustrate the LDOS plot for the (12,0) BNNT plus CLT of the top valence band between -1.07 to -0.93 eV and bottom conduction band between 1.00 to 1.14 eV. We can observe that the largest contribution to the top valence states in the regions under consideration comes from the N atoms of the BNNT (Fig. 4 (g), (i) and (k)), while for the conduction band region, the contributions are exclusive from the CLT molecule (Fig. 4 (h), (j) and (l)). These theoretical results may emphasize the nature of the π-π stacking interaction between CNT and CLT, which has already been suggested in the experimental results [32].

**Fig. 4** *Local density of states (LDOS) plot, with 95% probability of locating the electrons, for the (a) (8,0) CNT + CLT upper valence band (-0.39 to -0.27 eV),(b) (8,0) CNT + CLT lower conduction band (0.29 to 0.41 eV), (c) (10,0) CNT + CLT upper valence band (-0.42 to -0.29 eV),(d) (10,0) CNT + CLT lower conduction band (0.35 to 0.50 eV),(e) (12,0) CNT + CLT upper valence band (-0.28 to -0.20 eV),(f) (12,0) CNT + CLT lower conduction band (0.23 to 0.32 eV),(8,0) BNNT + CLT upper valence band (-1.09 to -0.97 eV); (h) (8,0) BNNT (8,0) + CLT lower conduction band (0.99 to 1.11 eV),(i) (10,0) BNNT + CLT upper valence band (-1.09 to -0.97 eV),(j) (10,0) BNNT + CLT lower conduction band (0.99 to 1.12 eV),(k) (12,0) BNNT + CLT upper valence band (-1.07 to -0.93 eV),(l) (12,0) BNNT + CLT lower conduction band (1.00 to 1.14 eV). The yellow, green, light blue and gray spheres represent the C, Cl, N and B atoms, respectively.*

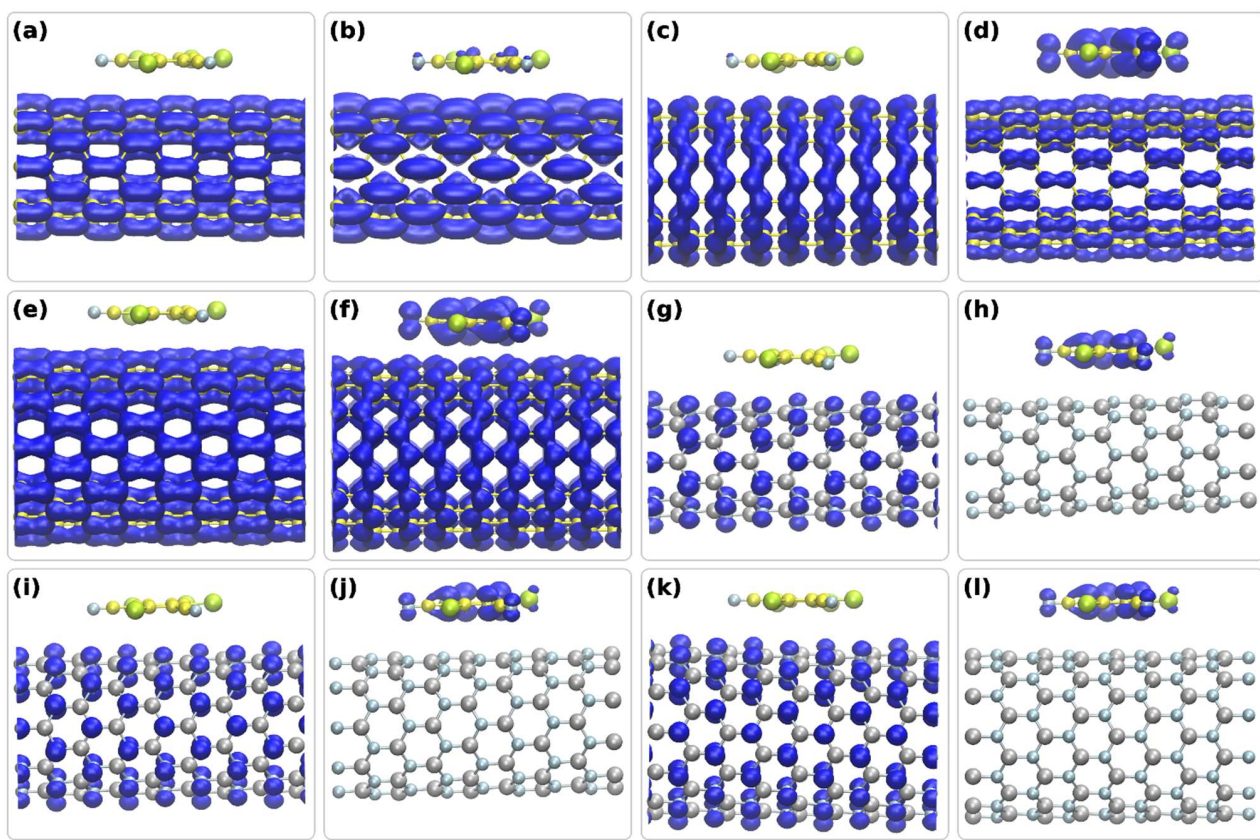

### 3.3 Interaction nature of the most relevant adsorption configuration for sensing applications

To better understand the differential interaction of CLT between these two nanotubes, we plotted the partial density of states (PDOS) of each system with respect to the contributions of the *p*-functions (Fig. 5). Since both nanotubes consist of alternating double bonds and CLT is an aromatic system, we expect a π-π stacking interaction between the nanotubes and CLT. In Fig. 5 (a), (b) and (c), we see the three PDOS for each CNT chirality, corresponding to the systems shown in Fig. 1 (a), (b) and (c), respectively. In the region between -5.0 to -2.5 eV, we observe a coupling between the states of the CNT C-2p functions and the C-2p and N-2p functions of the CLT molecule for the three CNTs considered.

The LDOS isosurfaces for the highest energy states in this region are depicted in Fig. 6 (a), (b), and (c). These plots make it evident that as the CNT radius increases, the spatial overlap between the CLT and CNT wavefunctions becomes more pronounced, with the (12,0) CNT showing the most significant overlap. This indicates a stronger π-π stacking interaction between the (12,0) CNT and the CLT than for the other nanotubes, which is related to nanotube radius. Specifically, as the nanotube radius increases, its reduced curvature allows for a more parallel alignment between the p-orbitals of the CNT and the CLT molecule [54]. This improved alignment enhances the subsequent π-orbital overlap, strengthening the π-π stacking interaction and explaining the increase in adsorption energy with increasing radius (Table 1).

The PDOS for the 2p states of the BNNTs and CLT atoms are shown in Fig. 5 (d)-(f), which correspond to the configurations in Fig. 1 (d)-(f). In the energy window between -2.5 to -2.0 eV, we observe a coupling between the 2p states of the B and N atoms of the BNNTs and the 2p states of the N and C atoms of the CLT molecule. In Fig. 6 (d)-(f), we plotted the LDOS in this energy window, and we observe the overlap between the isosurfaces around the BNNTs and the molecule (mainly on 99% probability) also increasing with the nanotube radius.

*Table 3 – Nanotube radius (r) and average bond length in each nanotube.*

| Systems | [a]$r$ (Å) | [b]$d$ (Å) |
|---|---|---|
| (8,0) CNT | 3.22 | 1.43 |
| (10,0) CNT | 3.98 | 1.43 |
| (12,0) CNT | 4.79 | 1.43 |
| (8,0) BNNT | 3.25 (3.32) | 1.46 |
| (10,0) BNNT | 4.05 (4.11) | 1.46 |
| (12,0) BNNT | 4.86 (4.90) | 1.46 |

[a] Values inside (outside) parenthesis represents the N-N (B-B) radius in BNNT.
[b] For CNT, bond length are for C-C distances, while for BNNT, they are B-N distances.

**Fig. 5** Partial density of states (PDOS) with respect to the p-functions contributions for the interacting systems with (a) (8,0) CNT + CLT, (b) (10,0) CNT + CLT, (c) (12,0) CNT + CLT, (d) (8,0) BNNT + CLT, (e) (10,0) BNNT + CLT and (f) (12,0) BNNT + CLT. The Femi level was shifted to zero.

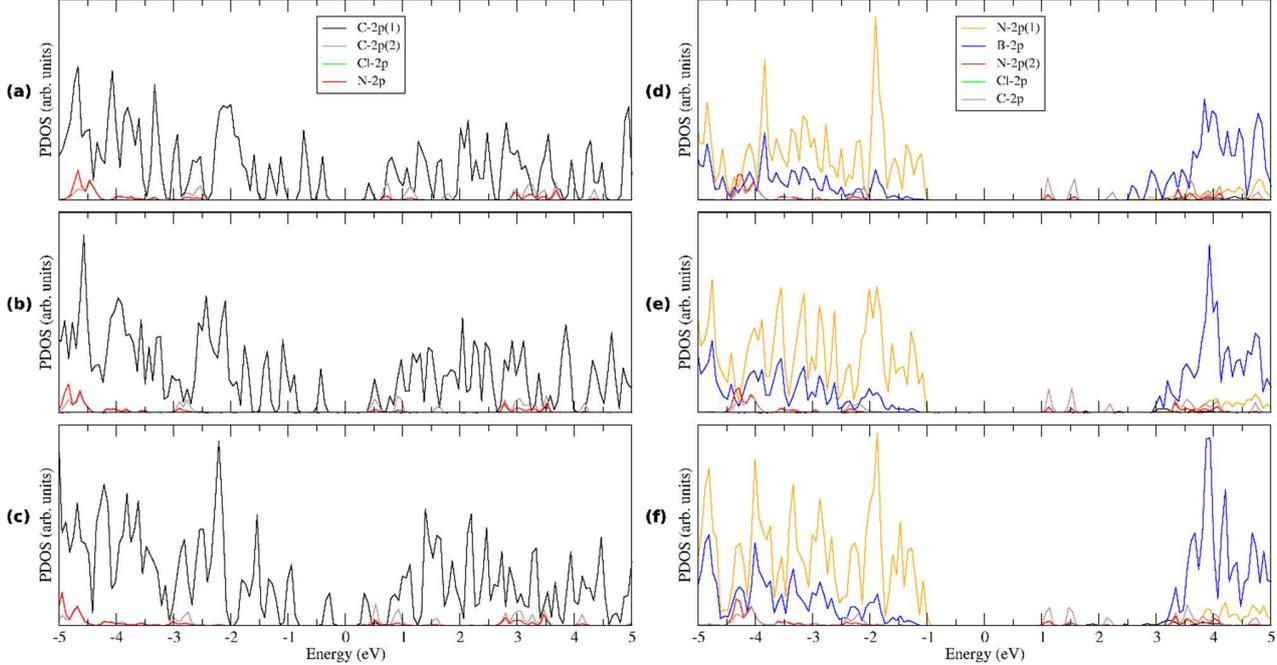

Directly comparing the LDOS with CLT adsorbed on the walls of CNTs (Fig. 6 (a)-(c)) and BNNTs (Fig. 6

(d)-(f), we observed less overlap between the isosurfaces around CLT and BNNT compared to the CNT systems, even for (12,0) BNNT. Since the radii of BNNT and CNT are very similar, as are the bond lengths (Table 3), the difference in curvature cannot explain the lower π-π interaction between these systems. So, to explain this effect, we need to compare the same electron density for each state, but with a reduced probability of finding the electron to 95% (Fig. 6 left) instead of 99% (Fig. 6 right). When we do this, we see that the isosurface of BNNT is mainly on N atoms, while for CNT each C atom contributes equally to the isosurface. Thus, the lower π–π interaction in BNNTs can be attributed to the lower electron density in their π orbitals, which are primarily derived from N-2p orbitals.

We also know that the B atom has an empty p orbital in its ground state, which can receive electron density from the N atom. However, we must not forget that the N atom is more electronegative than the B atom. Therefore, we would expect the electron density to be found in the N atoms rather than the B atoms, and this is exactly what happens here. So, since the electron density is more distributed across the CNT, this leads to a higher π-π stacking interaction. The lower electron density on these orbitals therefore explains the lower π-π stacking interaction between CLT and BNNT compared to CNT.

**Fig. 6** LDOS with 95% (left) and 99% (right) probability of finding the electron, for the following selected valence band states from Fig. : (a) (8,0) CNT + CLT (-2.87 to -2.45 eV), (b) (10,0) CNT + CLT (-3.02 to -2.63 eV), (c) (12,0) CNT + CLT (-3.09 to -2.68 eV), (d) (8,0) **BNNT + CLT (-2.41 to -2.02 eV),** (e) (10,0) BNNT + CLT (-2.47 to -1.97 eV) and (f) (12,0) BNNT + CLT (-2.50 to -2.07 eV). The yellow, green, light blue and gray spheres represent the C, Cl, N and B atoms, respectively.

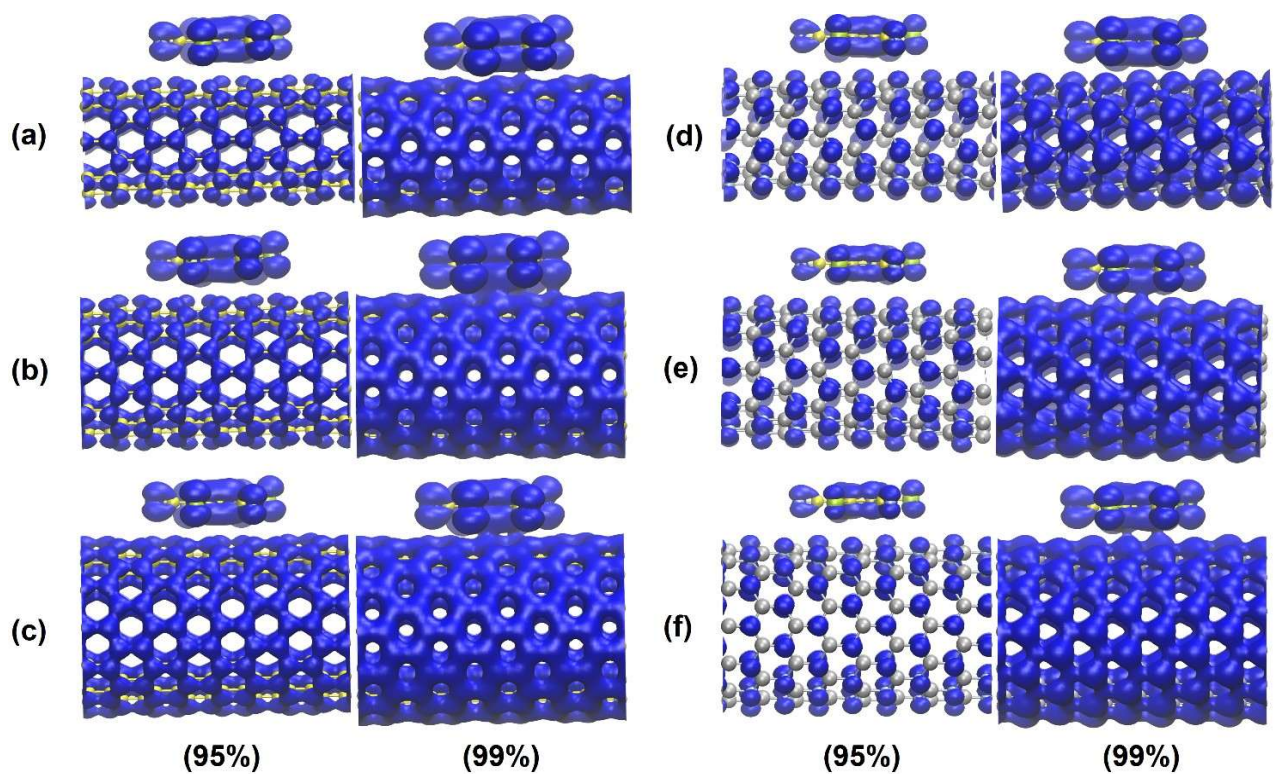

The increased strength of the π-π stacking interaction with increasing radius, especially for the CNTs, is perfectly corroborated and explained by the analysis of the global reactivity descriptors. According to the conceptual DFT framework, charge transfer spontaneously occurs from the species with a higher chemical potential (less negative) to the one with a lower chemical potential (more negative). In all cases, the chemical potential of the pristine nanotubes is higher than that of the CLT molecule (Table 4). This confirms a charge transfer from the nanotube to the CLT molecule. The consequence of this electron donation is a decrease in the nanotube's chemical potential, which becomes more negative after the interaction as its remaining electrons become more tightly bound. The data in Table 4 precisely confirm this effect: the μ of the (8,0) CNT, for instance, shifts from -4.52 eV to -4.74 eV. The magnitude of this change is a direct indicator of the interaction's intensity: for the larger-radius CNTs, (10,0) and (12,0), the change in μ is drastically greater than for the (8,0), aligning perfectly with the idea that a better π-π overlap facilitates a much more significant charge transfer.

The chemical hardness (η) explains why the CNTs are more efficient in this electron-sharing interaction. According to the Hard-Soft Acid-Base (HSAB) principle [55,56], hardness measures a species' resistance to the deformation of its electron cloud (polarizability). The CLT molecule (η = 4.0) and the BNNTs (η ≈ 2.3-2.5) are chemically "hard" species, meaning they are not easily polarized. The CNTs (η ≈ 1.1-1.4), on the other hand, are considerably "softer." The π-π stacking interaction fundamentally benefits from the mutual polarizability of the π-electron clouds. As CNTs are softer and more polarizable than BNNTs, they are intrinsically better suited to establish a strong π-π stacking interaction with CLT, which facilitates the charge transfer. The greater hardness of the BNNTs, on the other hand, limits this electronic overlap, resulting in the weaker interaction observed.

Table 4 – Chemical potential (μ), chemical hardness (η) and electrophilicity index (ω) for each system. Values in parenthesis represent the interacting system.

| Systems | μ | η | ω |
|---|---|---|---|
| (8,0) CNT | -4.52 (-4.74) | 1.39 (1.39) | 7.36 (8.08) |
| (10,0) CNT | -4.50 (-5.56) | 1.35 (1.29) | 7.50 (8.08) |
| (12,0) CNT | -4.48 (-5.51) | 1.09 (1.06) | 9.16 (9.60) |
| (8,0) BNNT | -4.63 (-4.97) | 2.33 (2.38) | 4.60 (5.18) |
| (10,0) BNNT | -4.43 (-4.87) | 2.45 (2.34) | 3.99 (5.08) |
| (12,0) BNNT | -4.30 (-4.81) | 2.45 (2.25) | 3.77 (5.15) |
| CLT | -5.64 | 4.00 | 3.97 |

Finally, the electrophilicity index (ω) measures the ability of a species to accept electrons and stabilize its energy upon doing so, i. e., it quantifies the energetic outcome of the interaction. The pristine CNTs are inherently more electrophilic (higher ω) than the BNNTs. After the interaction, ω increases for all systems, indicating that the nanotube-CLT system as a whole is a better electron acceptor than the isolated nanotube. This suggests the formation of an energetically stable system. For the BNNTs, the increase in ω is more pronounced than for the CNTs, suggesting that although the interaction is weaker, it induces a more drastic electronic perturbation in the BNNT system. This directly explains why their electronic properties, such as the band gap and work function, also exhibit more significant changes upon interaction compared to the CNTs.

### 3.4 Stability in aqueous solution through molecular dynamic simulation

The stability of these systems on water solution were investigated through molecular dynamics simulations. The total energy plots for the 100 ns production run (Fig. 7 (a) and 7 (b)) confirms that both systems reached an equilibrated state. Throughout the entire simulation, the energy fluctuates consistently around a well-defined average value, exhibiting no systematic drift or abrupt variations. This behavior confirms that the systems were properly equilibrated before the production step and maintained their dynamic stability and structural integrity during the entire simulation time.

As expected, the most stable conformation found in the DFT calculation was maintained throughout the MD simulations, with the molecules preferentially adsorbed on the tube surface with their aromatic ring parallel to the tube's surface. This can also be verified from the radial distribution function (RDF) in Fig. 7 (c), in which the most probable distance (first peak of the RDFs) between the carbons of the aromatic CLT ring and each nanotube surface is 3.85 Å and 3.65 Å for CNT and BNNT, respectively. The coordination number were approximately 2 in both cases, which means that each carbon of the aromatic CLT ring is close to at least 2 nanotube atoms at this distance. This proves that the preferred conformation of these molecules is parallel to the tube surface, as we can see on the last frame of the trajectory in Fig. 7 (d), (e), (f) and (g).

The higher values of g(r) also show that the CLT has a higher affinity to the nanotube surface than to water, indicating that both nanotubes are able to adsorb and remove this molecule from the environment, which is confirmed by the experimental result with CNT as stationary phase for solid phase extraction [32]. The second peak of the RDF (Fig. 7 (c)) represents the correlation of the CLT molecule with the opposing nanotube surface. Consequently, the distance between these first and second peaks corresponds to the nanotube's diameter. The robustness of this adsorption is evidenced by the absence of any desorption events throughout the entire MD simulation. The molecule remained stably anchored to the surface, suggesting that the interaction energy is significant and the timescale required for desorption far exceeds the simulation time, which agrees with our estimated recovery time. Finally, the consistency of these findings is supported by the overall stability of the system. The kinetic energy and temperature curves (Fig. 7 (h) and (i), respectively) fluctuate consistently around their equilibrium plateaus, confirming that the systems are at thermal equilibrium and are dynamically stable at room temperature.

**Fig. 7** Evolution of the total energy during the MD production run for the (10,0) (a) CNT and (b) BNNT interacting with CLT molecules. (c) Radial distribution function for the correlations between the carbon from the aromatic rings of CLT with each nanotube atoms (excluding the terminal C-H, B-H and N-H bonds). Representation of the last frame of the production trajectory for CNT + CLT in (d) side and (e) front view, and BNNT + CLT in (f) side and (g) front view, with all water molecules suppressed to better a visualization of the system, where the yellow, gray, light blue and green spheres represent the C, B, N and Cl atoms, respectively. Moving average plots of (h) temperature and (i) kinetic energy from the MD production step for the CNT + CLT (green curves) and BNNT + CLT (blue curves) systems.

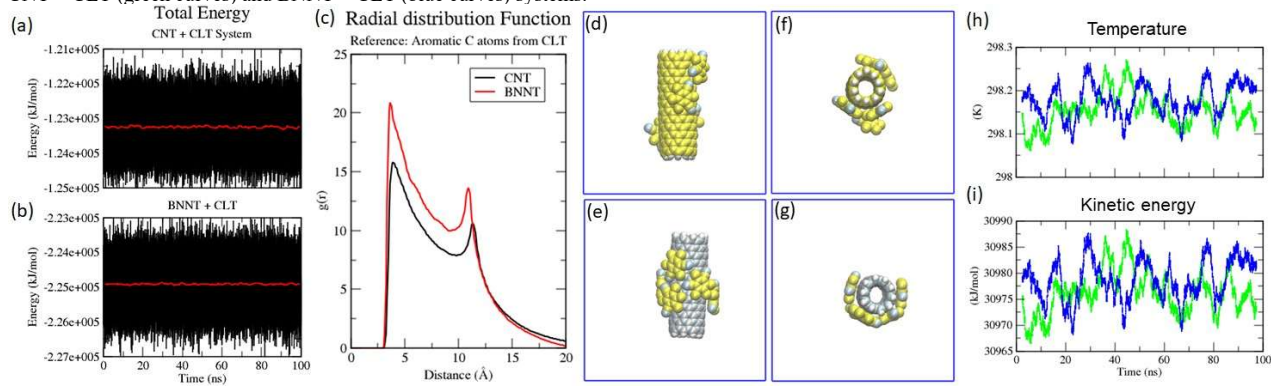

## 4. Conclusions

The interaction of the chlorothalonil molecule with CNT and BNNT was investigated through electronic structure calculations based on DFT and molecular dynamics simulations. The results showed that the adsorption energies were moderate in both systems. Thus, due to the moderate adsorption process, these nanotubes can achieve a reversible cycle of adsorption and desorption in applications for sensors. The molecular dynamics simulations also confirmed the thermal and dynamic stability of both systems in aqueous solution, with the molecule adsorbed onto both nanotubes surface at the same preferential position predicted by DFT calculations.

Our results show that both nanotubes are able to remove the chlorothalonil molecule from the environment, however, only the BNNTs are suitable for the detection of chlorothalonil because they have a shorter recovery time compared to CNTs and their band gap is drastically reduced due to the appearance of new defect energy levels in the gap region, while the CNT band gap remains practically unchanged. As a result, BNNTs respond better to CLT detection due to the change in electrical conductivity and better electronic sensitivity, making them a good candidate for the detection and removal of the pollutant chlorothalonil from contaminated environments.

CNTs are not well-suited to work-function-based applications, as their work-function does not significantly change upon molecule adsorption. For work-function-based applications using BNNTs, their radius might be well-controlled, as their work function response significantly diminishes as the tube diameter increases. Although this introduces a limitation (requiring careful control over the BNNT diameter), BNNTs still remain better candidates than CNTs for electronic sensing of this molecule, while also allowing for some tunability in sensor design.

The interaction of chlorotalonil with the surfaces of both CNTs and BNNTs increases as the radius grows.

This trend is driven by a more effective π-π stacking interaction on the less curved and more planar surfaces of larger nanotubes. This geometric argument, however, is deeply related to the fundamental electronic properties of these systems, as revealed by the global reactivity descriptors. The superior adsorption energies of CNTs is explained by their greater polarizability (lower hardness, η), which makes them intrinsically better suited for π-π stacking interaction. This, in turn, facilitates a more significant charge transfer from the nanotube to the molecule, a phenomenon unequivocally confirmed by the calculated shifts in the chemical potential (μ). While the interaction is weaker for BNNTs, their electronic structure is more profoundly impacted by the adsorption, as indicated by the pronounced increase in their electrophilicity index (ω) upon interaction.

Since the radius of a (n,0) nanotube is directly related to the Hamada index (n), it is expected that nanotubes with higher n values will exhibit higher adsorption energies with this molecule. However, a physically intuitive trend of stronger adsorption on flatter surfaces is governed by a nuanced interplay of polarizability and charge transfer. Even as the interaction energy grows with the radius, the nature of the interaction remains a physisorption process. Therefore, we can confidently conclude that for even larger nanotubes, the interaction will not evolve into chemisorption but will instead approach an upper energy limit characteristic of a strong physical adsorption.

**Acknowledgments**

The authors are grateful for the financial support provided by CAPES (Finance Code 001).
Francisco Gleidson de Sousa Ferreira and Caio V. C. Ribeiro da Silva acknowledge CAPES for the fellowship.
Kleuton A. L. Lima acknowledges the financial support from FAPESP (Grant No. 2024/21870-8).

**Author Contributions**

**Francisco Gleidson de S. Ferreira**: Data Curation, Formal Analysis, Investigation, Software, Visualization, Writing – original draft. **Caio V. C. Ribeiro da Silva**: Data Curation, Formal Analysis, Investigation, Software, Writing – review & editing. **Silvete Guerini:** Writing – review & editing. **Kleuton A. L. Lima:** Writing – review & editing. **Douglas Soares Galvão:** Writing – review & editing. **José Milton Elias de Matos:** Writing – review & editing. **Alexandre Araujo de Souza**: Conceptualization, Methodology, Resources, Supervision. **Francisco Gleidson de S. Ferreira** and **Caio V. C. Ribeiro da Silva** contributed equally to the work.

**Declaration of Competing Interest**
The authors declare that they have no known competing financial interests or personal relationships that could have appeared to influence the work reported in this paper.